# Subgrid-scale modeling for microbubble generation amid colliding water surfaces

W. H. R. Chan, J. Urzay, and P. Moin
(Center for Turbulence Research, Stanford University, USA)

**ABSTRACT**

The generation of microbubbles upon the collision and interaction of liquid bodies in a gaseous environment is a ubiquitous process in two-phase flows, including large-scale phenomena like ship wakes (Trevorrow *et al.*, 1994; Reed and Milgram, 2002; Zhang *et al.*, 2004; Stanic *et al.*, 2009), breaking waves (Melville, 1996; Deane and Stokes, 2002) and rain showers (Blanchard and Woodcock, 1957; Prosperetti and Oğuz, 1993). These collision and interaction events involve the relative approach of pairs of liquid-gas interfaces. As these interfaces approach, the smallest length scales of the system are dynamically reduced. This evolving disparity in length scales is numerically challenging to resolve without the employment of subgrid-scale (SGS) impact and breakup models. In this study, a physics-based impact and breakup model for the generation of these microbubbles is developed and implemented.

The objectives of this study are to develop a computational algorithm that identifies interface collision events that contribute to the formation of microbubbles, to formulate a physics-based breakup model that predicts the distribution of microbubble sizes using the characteristics of the originating gas film, and to integrate these modules into a two-phase flow solver that accurately captures the effects of bubbles of all sizes.

In these proceedings, an SGS model suitable for the aforementioned problems is proposed, and the steps involved in implementing the proposed SGS model in a macroscale flow solver are outlined. Two aspects of the development of this SGS model are then discussed in detail. First, the formulation and implementation of the first step of the SGS model, the collision detection algorithm, is detailed. Identification at the correct location and time of collision events that contribute to the formation of microbubbles is crucial to the appropriate activation of the SGS model in the macroscale flow solver. Second, preliminary findings of a numerical investigation intended to shed light on breakup processes in turbulent two-phase flows are presented. A parameter study is performed for a breaking wave, varying the ratio of some of the characteristic energy-containing scales of the system, so as to permit the resolution of a class of bubbles – the sub-Hinze scale bubbles (Hinze, 1955; Deane and Stokes, 2002) – that is typically infeasible to resolve in numerical simulations of the energetic breaking waves targeted by the proposed SGS model.

**INTRODUCTION**

**The need for an SGS model**

As two liquid bodies approach each other, such as when liquid droplets impinge on liquid pools, small amounts of gas become entrapped in thin films between the liquid surfaces. In certain physical regimes (see, for example, Fig. 2), flow instabilities cause the gas films to break up into smaller tiny bubbles that are later dispersed. For example, around moving ships, air bubbles of a broad range of sizes are often entrained due to boundary layers and stern waves, forming an elongated wake that lasts for several kilometers downstream (Reed and Milgram, 2002). Small microbubbles rise to the surface slowly and persist due to their low terminal velocity, and contribute to the physical impact of these long-reaching wakes. It is thus important to accurately capture the presence and behavior of these thin films and microbubbles in simulations of these systems. Yet, it is numerically challenging to resolve these thin films and microbubbles together with the largest scales of the system, such as the ship length. Simultaneous simulation of these disparate scales with feasible computational cost demands the development of SGS models for the small-scale features and resolution of the remaining large-scale features using large eddy simulation (LES).

**Abstractions of possible model problems for the SGS model**

The development of an SGS model for collision and rupture events requires the abstraction of a simple and localized canonical problem with a lower solution cost or solutions that can be tabulated for future reference. This is not a straightforward task for complex turbulent flows in two-phase environments due to the wide variation in the scales of the participating flow structures. However, from geometrical considerations, the elementary microscale process of these collisions almost always involves two curved liquid surfaces approaching and locally impacting each other, as depicted in the top left and top center panels of Fig. 1. An adequate model problem, then, involves the approach and impact of two liquid quadric surfaces with arbitrary curvatures, as depicted in the center panel of Fig. 1. Unfortunately, this problem remains difficult to comprehensively characterize to date, and experiments involving this generalized geometry are scarce. Instead, a simpler model problem, depicted in the bottom panel of Fig. 1, is investigated in this work: the impact of a spherical liquid drop on a deep liquid pool with a flat surface. Headway in this problem has been made analytically (Thoroddsen *et al.*, 2005; Mani *et al.*, 2010; Duchemin and Josserand, 2011; Hicks and Purvis, 2011; Mandre and Brenner, 2012; Bouwhuis *et al.*, 2012; Hendrix *et al.*, 2016) and experimentally (see references in Fig. 2), making the problem a reasonably well-understood base geometry for solution and tabulation, although several details like the relative importance of various forces at very small interface separations and the precise mechanisms of rupture remain topics of research.

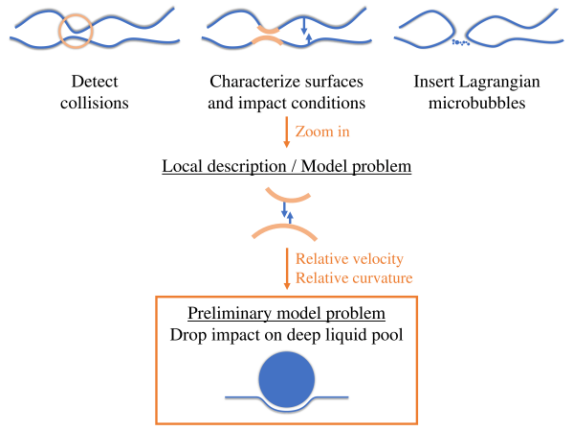

**Figure 1:** Schematic illustrating the abstraction of a model problem for microbubble generation upon the impact of two liquid surfaces, as well as the role of the model problem in the SGS model (to be discussed in the next section).

Consider a drop with speed $U$ and diameter $D$ falling onto a liquid-gas interface with surface tension $\sigma$. Nondimensionalization of the Navier-Stokes equations describing the motion of the drop, as well as the motion of the gas separating the drop and the liquid surface, along with the coupling conditions at the interfaces between the two phases, yields several dimensionless quantities. These are, namely, the density ratio

$$\frac{\rho_l}{\rho_g} = \frac{\text{liquid density}}{\text{gas density}}, \quad (1)$$

the viscosity ratio

$$\frac{\mu_l}{\mu_g} = \frac{\text{liquid viscosity}}{\text{gas viscosity}}, \quad (2)$$

the Weber number (We)

$$\text{We} = \frac{\rho_l U^2 D}{\sigma} = \frac{\text{inertial forces}}{\text{capillary forces}}, \quad (3)$$

and the Stokes number (St)

$$\text{St} = \frac{\rho_l U D}{\mu_g} = \frac{\text{drop relaxation time}}{\text{flow time}}. \quad (4)$$

In a typical simulation, the gas and liquid materials are usually predetermined (e.g., air and water in oceans), thereby fixing the density and viscosity ratios of the system throughout the computation. One then expects the phenomenology of the bubbles produced from film rupture, such as the occurrence of microbubbles, in a given collision event to be solely a function of We and St. To support this hypothesis, a survey was conducted of various experiments of water drops impinging on deep water pools with planar surfaces under atmospheric conditions, and some of these data are plotted in Fig. 2. The plot suggests that the resultant bubble phenomenology can be predicted, and thus tabulated, based on We and St. For example, the region $10 < \text{We} < 100$ and $6 \times 10^4 < \text{St} < 3 \times 10^5$ appears to describe the formation of microbubbles. An analysis of the corresponding visual data suggests many of these microbubbles are formed from the rupturing of a hemispherical air film trapped between the drop and the surface (Sigler and Mesler, 1990; Mills *et al.*, 2012), a distinct film phenomenology from the other cases.

**Outline of the paper**

Several questions immediately arise following the preliminary analysis of the model problem selected above. For instance, how does one determine when and where these collision events occur in a macroscale flow solver? What dimensionless parameters correspond to these events, if macroscale flow solvers are inherently poorly disposed to resolving the thin films resulting from these impact events? In addition, is the film rupture mechanism implied by the selected model problem a true reflection of how small features in turbulent two-phase flows with colliding surfaces are generated? In this study, an attempt is made to address some of these issues. A collision detection algorithm intended to identify these events is developed and implemented, and a numerical investigation of the mechanisms of the generation of small features in a breaking wave is embarked upon. Before these issues are addressed, the proposed SGS model alluded to in Fig. 1 is revisited, and the ideas introduced in the top panels of Fig. 1 are developed in more detail.

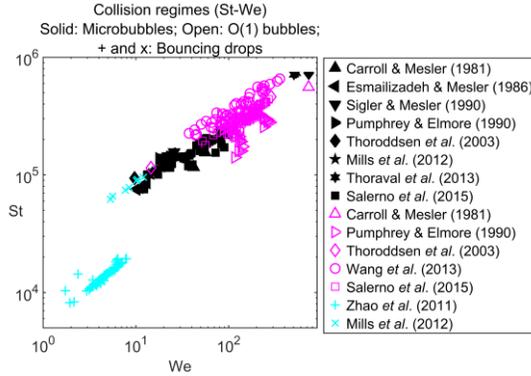

**Figure 2:** Experimental data of We and St associated with water drops impinging on deep water pools under atmospheric conditions at various collision regimes. The solid symbols indicate drops that produce microbubbles, while the open symbols indicate drops that produce only a handful of macrobubbles and the crosses indicate drops that bounce without producing microbubbles.

## SGS MODEL FORMULATION

In the introduction, a model problem amenable to solution tabulation was identified, making an SGS model for microbubble generation feasible. As suggested in the top panels of Fig. 1, two other procedures then remain for successful implementation in a macroscale flow solver: the activation of the SGS model, and the insertion of features generated by the SGS model. In light of this, the following sequential subproblems are proposed for the SGS model: i) the detection of imminent collisions by exhaustive search throughout the macroscale simulation domain, ii) the characterization of the surface geometry and relative impact velocity of the approaching interfaces, iii) the calculation of the breakup dynamics of the entrapped gas film, and iv) the insertion and transport of Lagrangian microbubbles in the macroscale flow solver upon eventual collision of the interfaces. These subproblems are described in greater detail in Fig. 3 and in the subsequent discussion.

### Identifying and characterizing impact events

In a true direct numerical simulation (DNS) of turbulent two-phase flows, all the features in the flow, including the impinging surfaces, the entrapped gas films and the resultant microbubbles, as well as the key physical phenomena associated with their formation, evolution and demise, need to be well-resolved, as depicted on the left of Fig. 4. This is a stringent requirement on the grid resolution. Better intuition of this requirement can be obtained by considering the grid Weber number, along with the Weber numbers associated with turbulent velocity fluctuations at the scales of the aforementioned features. Define the Weber number associated with velocity fluctuations of magnitude $u_n$ at a characteristic length scale $l_n$ as

$$\text{We}_n = \frac{\rho_l u_n^2 l_n}{\sigma}. \tag{5}$$

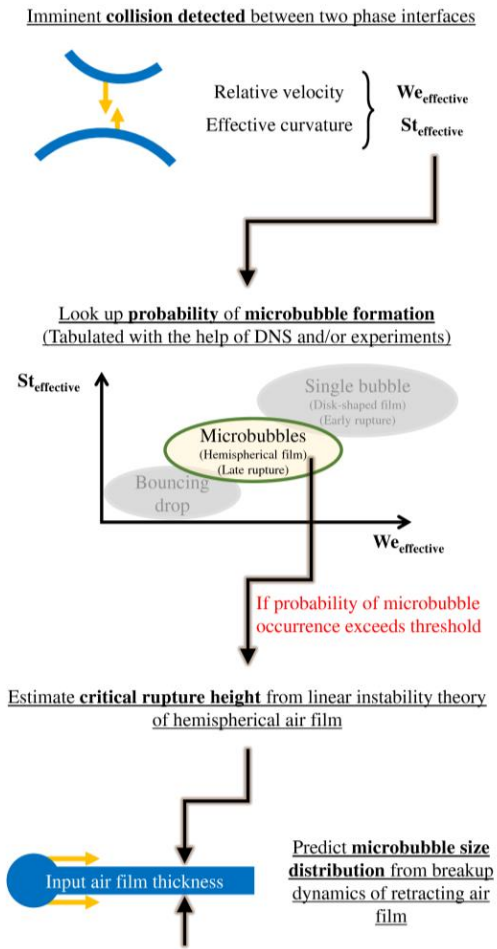

**Figure 3:** Detailed schematic of SGS model for the generation of microbubbles upon the impact of two water-air interfaces.

If $l_n$ is equal to the grid resolution $\Delta$, and $u_\Delta$ is the corresponding characteristic velocity fluctuation magnitude at this scale, then

$$\text{We}_\Delta = \frac{\rho_l u_\Delta^2 \Delta}{\sigma} \quad (6)$$

represents the grid Weber number. Many recent 3D simulations of hydraulic jumps and breaking waves on structured grids (Wang *et al.*, 2016; Mortazavi *et al.*, 2016; Deike *et al.*, 2016) have been able to achieve $\text{We}_\Delta$ on the order of 0.1 to 1, albeit at the expense of high computational cost. This is sufficient to resolve bubbles that break up due to the action of turbulent fluctuations ($\text{We}_n \gtrsim 1$ where the length scale is based on the bubble radius), but not most of the bubbles below the Hinze scale that are formed due to the action of capillary forces ($\text{We}_n \ll 1$). The Hinze scale

$$l_n = l_\text{H} \sim \frac{\sigma}{\rho_l u_n^2} = \frac{\sigma}{\rho_l u_\text{H}^2} \quad (7)$$

corresponds to $\text{We}_n \sim 1$ and is discussed in greater detail in a subsequent section. It is assumed here that the Hinze scale is larger than the Kolmogorov scale. When this is true, the inertial subrange scalings remain valid for velocity fluctuations and length scales just below the Hinze scale, to the extent that local isotropy at the length scale being considered can be assumed. These considerations are also further discussed later. As one considers length scales increasingly smaller than the Hinze scale (but above the Kolmogorov scale), the effects of surface tension become increasingly dominant relative to the effects of the turbulent fluctuations, since the corresponding $u_n$ and thus $\text{We}_n$ at this scale are decreased. The dominance of capillary effects implies the increase in the relative importance of capillary-driven motion like thin film retraction. Such motion is typically associated with a Weber number of order 1 (as evident, for example, from the expression for the Taylor-Culick speed (Taylor, 1959; Culick, 1960; Mirjalili and Mani, 2018)), which suggests that a thinning feature should be associated with a higher capillary-driven velocity $u_c$. A true DNS, then, requires a sufficiently small $\text{We}_\Delta$ such that these capillary-driven velocities $u_c$ dominate the turbulent velocity fluctuations $u_n$ at the smallest resolved scales and are well captured. Consider an affordable LES with $\text{We}_\Delta \gtrsim 1$ such that all the sub-Hinze scale microbubbles, as well as other sub-Hinze scale phenomena, are treated with an SGS model, as depicted on the right of Fig. 4. In this simulation, it is unlikely that either the microbubbles or their precursor gas films are well-resolved. The SGS model will then have to be activated before the entrapped gas film becomes too thin and the mesh can no longer resolve the dynamics of the film. Note that the curvature of the impacting interfaces is still expected to be resolved for high-aspect-ratio films. Low-aspect-ratio films correspond to under-resolved or subgrid interface corrugations, which will need to undergo further modeling.

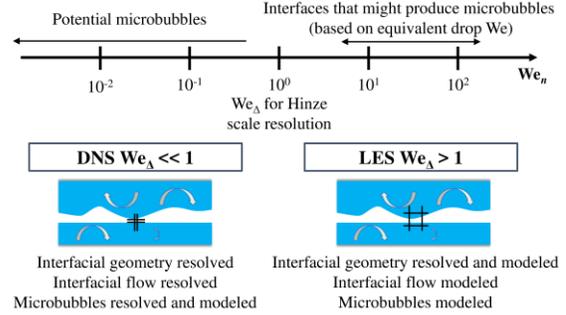

**Figure 4:** Schematic comparing relevant length scales and resolution of features in typical DNS and LES of turbulent two-phase flows. Since the formation of microbubbles due directly to the impact of two arbitrarily curved interfaces has not been well-characterized, the top panel of the schematic explicitly refers to the equivalent drop We in the drop-pool model problem, where microbubbles at least 20 to 50 times smaller than the impinging drop have been visually observed (see references with solid symbols in Fig. 2). The black lines in the two sub-schematics depict the effective grid size relative to the illustrated surfaces.

The transition from the macroscale flow solver to the microscale model problem has to be managed delicately. On one hand, as the film between a pair of approaching liquid bodies becomes thin, the macroscale solver becomes progressively ill-conditioned to handle the film. On the other hand, the inputs to the model problem should involve quantities as close to impact as possible. Experimentally, most quantities are reported at the time just prior to impact since it is usually infeasible to report or measure the time evolution of these quantities. This evolution may be more accessible in some theoretical and numerical approaches, but the ultimate tabulation of solutions has to be based on a single set of dimensionless input parameters — most likely involving the quantities just before impact for compatibility with experiments — in order to facilitate its implementation. This would enable a grid-independent SGS model that only requires knowledge of the impact conditions, or at least a robust estimate of them. As such, the detection of imminent collision events in the macroscale flow solver should be delayed to the instant just before the mesh can no longer resolve the film, in order to ensure maximum correspondence between the quantities measured at the moment of transition in the macroscale flow solver and the impact parameters assumed by the microscale model problem. The film thickness at this point could possibly be on the order of a couple of grid cells.

In order to bridge the macroscale flow solver and the microscale model problem, the following methodology is proposed. At every time step in the macroscale flow solver, an exhaustive search is employed throughout the simulation domain to locate all interface pairs that are potentially about to collide. The gas film between each of these pairs has to be on the verge of becoming poorly resolved, and the relative motion and orientation of the interfaces should result in imminent coalescence. The geometric parameters of the film trapped between each pair of potentially colliding interfaces, such as the surface curvatures and the relative approach velocity of the interfaces, are then quantified using the velocity and phase fields of the flow solver — in a volume-of-fluid solver, for example, the latter would refer to the volume fraction field. Finally, the dimensionless parameters that arise from these quantities are adopted as the input parameters of the microscale model problem. Note that the two arbitrary curvatures measured in the macroscale flow solver for each potential collision will need to be mapped to an effective drop curvature for the model problem, as will be discussed in a subsequent section.

**Solving the model problem**

Once the effective impact We and St are computed for a particular imminent collision, a lookup table constructed with the aid of DNS (e.g., Mirjalili and Mani, 2018) and experiments is then consulted to determine the physical regime in which the detected potential collision resides. The lookup table is expected to take a form similar to Fig. 2, and a detailed parameter sweep using DNS, or interpolation by kriging or deep learning methods, may help to increase the fidelity of the table. It has been observed experimentally (Mills *et al.*, 2012) that on separate runs with the same parameters, bubbles of different phenomenology can occur. For example, one run with a particular We and St might produce microbubbles, but another with the same We and St might instead produce a single bubble. The lookup table is thus best formulated in terms of probabilities, as suggested in the second panel of Fig. 3.

If the effective We and St suggest that microbubbles are likely to be formed, then one will proceed to estimate the size distribution of the generated microbubbles based on the breakup dynamics of a retracting gas film with a finite boundary. The radial extent of the film is expected to be comparable to the inverse of the effective curvature of the approaching interfaces, leaving the thickness of the trapped gas film as the only input parameter to this model subproblem. If the dynamics of the rupture event that created the finite trapped gas film are dominated by van der Waals attractive forces, then the initial film thickness is typically on the order of 100 nanometers where van der Waals intermolecular and capillary forces balance each other (Baldessari *et al.*, 2007; Kaur *et al.*, 2009). Investigations are ongoing to determine if the initial film thickness can be modeled more precisely and if it is a strong function of the collision We and St. Work is also in progress (Mirjalili and Mani, 2018) to compute the microbubble size distribution from the retraction and breakup of finite gas films of various thicknesses. Future iterations of the SGS model will need to consider the possibility of coupling between nearly-simultaneous and closely-spaced (within the effective radius of curvature) collisions.

The final step of the SGS model involves the insertion of the predicted microbubbles at appropriate locations and velocities into the flow solver once the collision has been ascertained to have occurred. Since these microbubbles are unlikely to be supported by the solver mesh, a Lagrangian treatment should be used for their transport, as with other under-resolved features in the flow solver. The corresponding gaseous mass and momentum entrapped in the microbubble phase have to be removed from the relevant Eulerian fields to ensure conservation. Breakup and coalescence models will be necessary for these Lagrangian particles. Additional comments on several aspects of the proposed SGS model are provided in Chan *et al.* (2016) and Chan *et al.* (2017).

**COLLISION DETECTION ALGORITHM**

**Description of the algorithm**

Potential collision events need to be detected at the right place and the right time for appropriate activation of the SGS model. As alluded to earlier, it is crucial that a potential collision be detected immediately before resolution of the entrapped gas film is lost, even if the eventual collision does not take place in the next computational time step. In this algorithm, the detection of any potential collision between approaching interfaces is triggered once the interfaces come within a predetermined number of grid-cell widths of each other. Here, a heuristic argument is attempted to justify the choice of this requirement, and a specific example is highlighted.

Spatial gradients are crucial to solutions of the semi-discrete Navier-Stokes equations, as well as the accompanying advection equation for the phase field in an interface-capturing method. While the subsequent discussion can be generalized to different interface-capturing methods and numerical schemes, this work specifically considers a volume-of-fluid (VoF) scheme implemented in a node-centered finite volume code (e.g., Kim *et al.*, 2014). In such a solver, finite volume operators (e.g., Ham *et al.*, 2006) involving the unknowns at each node, as well as its neighbors, are used to compute these gradients. In the code cited above, second-order accurate operators are used. Hence, the stencil of the operators only involves the neighboring layer of nodes. Application of these operators to the volume fraction field is performed in the computation of the normal vectors and mean curvature at every node associated with an interface. These normal vectors and curvatures are necessary to solve the conservation and VoF advection equations, as well as to characterize the inputs to the model problem in the SGS model. Thus, potential collisions between two interfaces must be detected before the interfaces become so close that the orientations of their normal vectors influence each other. In the code cited above, this would demand triggering the collision detection once the interfaces come within two grid-cell widths of each other. Codes with higher orders of accuracy would accordingly require thresholds involving more grid-cell widths.

This numerical interference is demonstrated with a simple example illustrated in the schematic depicted in Fig. 5. Suppose there are two liquid bodies close to each other, as illustrated in the left subfigure of Fig. 5. The dotted lines correspond to the positions of the true physical interfaces, and the fill of the cells corresponds to the numerical distribution of liquid (filled) and gas (unfilled) in the computational cells. When the liquid bodies are sufficiently far apart, they are unaware of the presence of each other for the purpose of the computation of the normal vectors, since a single application of the gradient operator at every node extends only one node layer out. They are, of course, keenly aware of each other due to the incompressible nature of the solver. As such, the numerical distribution of liquid and gas is expected to closely match the positions of the true physical interfaces. Suppose the top liquid body is held fixed and the bottom liquid body is moved upwards, as illustrated by the dotted lines in the right subfigure of Fig. 5. Now, the distance between the interfaces is less than the average size of two grid cells, and the application of the gradient operator to a mesh node corresponding to one of the liquid bodies near its interface is expected to involve information in mesh nodes corresponding to the other liquid body. The eventual orientation of the normal vectors arising from this computation is likely to be such that numerical coalescence has taken place between the two bodies, and a possible numerical configuration is illustrated in the filled portions of the right subfigure of Fig. 5. To pre-empt this interference, it is proposed that potential collisions are detected once the interfaces come within two grid-cell widths of each other. This, however, implies that the eventual collision may not necessarily take place in the next time step if the interfaces are slowly approaching each other.

Because of the two-grid-cell widths requirement, the foundation of the collision detection algorithm, as implemented in the unstructured code cited above, rests on a node-of-node-of-node structure, where neighbors of neighbors of nodes are stored in a compressed sparse format for ease of lookup. An illustration of this structure is shown in Fig. 6. (In a code with higher-order operators, the corresponding structure required will involve more node layers.)

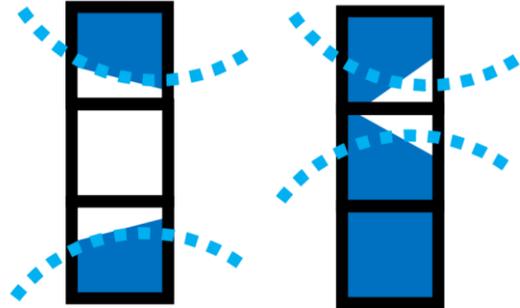

**Figure 5:** Schematic illustrating numerical interference between two liquid bodies in close proximity, in the context of the code cited in the main text. Interference occurs on the right, but not the left. Here, the black boxes refer to individual computational cells (median dual cells in a node-centered setting). The numerical interfaces are planar (here linear in a 2D projection) as is typical in a piecewise linear interface calculation (PLIC) VoF scheme. The significance of the lines and fills is explained in the main text.

The proposed collision detection algorithm proceeds as follows. (The ordering of the collision criteria in the description of the algorithm that follows differs from the actual algorithmic implementation for clarity and ease of illustration.) At every time step, nodes containing interfaces are looped over since a collision must involve two interfaces. For each interfacial node (hereafter denoted node A) with a location $\vec{x}_A$, a velocity $\vec{u}_A$ and a normal vector $\hat{n}_A$ that has not been tagged in a previous potential collision, any neighbor-of-neighbor B is selected for testing if

$$\frac{\vec{x}_B - \vec{x}_A}{|\vec{x}_B - \vec{x}_A|} \cdot \hat{n}_A \geq 0. \tag{8}$$

All eligible B in Fig. 6 have been shaded assuming the center node of the structure in the figure is A and $\hat{n}_A$ points upwards. Out of these nodes, only interfacial nodes permit a collision. A potential collision for an eligible A–B pair is deemed to be possible if all the mutual neighbors of A and B do not contain any liquid, if the two interfaces are facing each other

$$\hat{n}_A \cdot \hat{n}_B < 0, \qquad (9)$$

and if they are moving towards each other, implying

$$\frac{\vec{x}_B - \vec{x}_A}{|\vec{x}_B - \vec{x}_A|} \cdot \frac{\vec{u}_B - \vec{u}_A}{|\vec{u}_B - \vec{u}_A|} < 0, \qquad (10)$$

and

$$\frac{\vec{u}_B - \vec{u}_A}{|\vec{u}_B - \vec{u}_A|} \cdot \hat{n}_A < 0. \qquad (11)$$

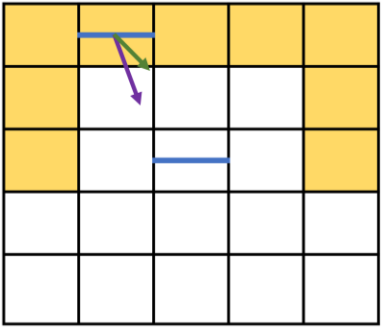

**Figure 6:** Median dual mesh illustration of node-of-node-of-node structure, in the context of the code cited in the main text. The cell currently being tested for collisions, denoted as node A in the main text, is the interfacial cell in the center of the illustration. The thick line splitting this cell represents a phase interface. Assume here that node A has an upward-pointing normal vector. The identity of the shaded cells is discussed in the main text. The other interfacial cell is denoted as node B in the main text. The two arrows could represent the normal vector of node B, the unit relative velocity vector of nodes A and B, or the unit relative displacement vector of the two nodes, and are meant to illustrate the tests in Eqs. (8) to (11).

Potential collisions involving interfaces that both have negative curvatures are rejected because they cannot be mapped to the drop-pool model problem discussed earlier and are physically unlikely to produce microbubbles. (This is discussed in more detail at the end of this section.)

After all nodes whose interfaces may undergo potential collisions have been identified, neighboring potential collisions are bundled so that a minimal set of new and unique potential collisions is obtained. An illustrated explanation of this bundling is presented in Fig. 7. Each bundled potential collision should have a unique location (center of mass) and effective curvature. This bundling is performed in two stages to ensure the parallelizability of the algorithm. Within a single processor, potential collisions are bundled using a flood-fill algorithm if they satisfy at least one of two conditions. First, the two potential collisions to be bundled should share a common mesh node. Second, the two potential collisions to be bundled should, between them, have a pair of nodes (one from each potential collision) whose separation is not more than two grid cells. For the second condition, potential collisions where the neighboring nodes have very different curvatures should not be bundled together even if they are close to each other. Across processors, potential collisions are then bundled using a union-find algorithm, where the union operation involves the same two conditions above with the same curvature restriction. These algorithms are similar to the "agglomerate" and "connect" algorithms discussed by Le Chenadec *et al.* (2014). Both algorithms have been modified so that the "interface parity" of each node involved in a potential collision is established and maintained: since each physical collision involves two interfaces, each potentially colliding node in the final bundled potential collision should be associated with exactly one of the two physical interfaces throughout the bundling process. This "parity" is illustrated using the difference in the shades of the colors of the crosses in the bottom subfigure of Fig. 7.

Even after this bundling process has taken place, each potential collision may still involve only a handful of mesh nodes. In order for the effective curvature of the potential collision to sample the curvatures of sufficiently many mesh nodes, the curvature field and any other quantities involved in the computation of the curvature are smoothed solely for the purpose of computing the effective curvature of each potential collision. The smoothing is performed for each node by averaging the relevant quantity over itself and its node neighbors. In this work, the final curvatures used to compute the effective curvature of the interface were chosen to be weighted averages of the original curvature and the smoothed curvature, so that the smoothing plays only a supplementary role in the final estimation of the interface curvature.

Without external intervention, the same pair of interfaces may trigger the collision detection algorithm at every time step thereafter since the interfaces satisfy the collision criteria but may not necessarily coalesce immediately. This is circumvented by the following procedure: once a potential collision is detected, mesh nodes in the neighborhood of the detected potential collision are used to generate two axis-aligned bounding boxes, one for each interface participating in the potential collision, that surround a reasonable portion of the associated interfaces. The bounding boxes are scaled such that they cover the extent of the larger of the two curvatures (i.e., the smaller of the two radii of curvature) of the interfaces. This allows the boxes to contain sufficient mesh nodes to block out this particular potential collision but not others. An illustration of these bounding boxes is shown in Fig. 8. The extents of these boxes are updated every time step using the average velocity of all the nodes within the boxes in a similar fashion to the forward Euler scheme. At every time step, if a node falls within the limits of both bounding boxes of any previously detected potential collision, then all new potential collisions involving the node are excluded from further consideration. This restriction is lifted once sufficient characteristic times for all the potential collisions associated with the node have elapsed, where the characteristic time of a potential collision is the ratio of the initial separation to the initial relative speed of the interfaces corresponding to that particular potential collision. Liquid bodies that have come into contact but not yet fully merged may continue to trigger the collision detection algorithm, so the number of characteristic times chosen needs to be sufficiently large to weed out these events.

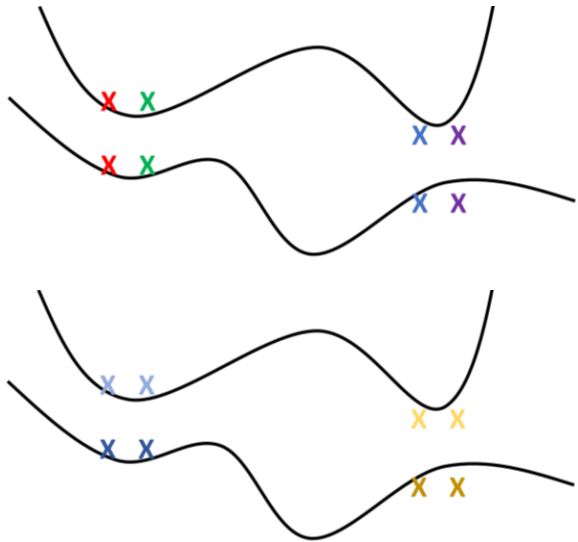

**Figure 7:** Schematics of bundling of new potential collisions between two approaching interfaces (marked with the thick black lines): before (top) and after (bottom). Suppose that the collision detection algorithm detects four different potential collisions, highlighted with four different colors in the top subfigure. Each potential collision involves two mesh nodes, and each mesh node is associated with exactly one of the two physical interfaces. The locations of the nodes where the algorithm was triggered are marked with crosses. Visual inspection suggests that the two potential collisions on the left should be bundled, as should the two on the right. The criteria and process of bundling are detailed in the main text. The bundling algorithm described in the main text ought to yield two bundled potential collisions, as suggested by the coloring of the crosses in the bottom subfigure. The top nodes and bottom nodes in each bundled potential collision should also be identified as belonging to two different interfaces, as suggested by the difference in the shades of the colors.

Each detected potential collision that fulfills the physical criteria outlined in the SGS model should eventually result in a bubble insertion event if the interfaces eventually do collide. The model does not, however, offer a clear choice for the insertion time. It is currently proposed that the bubbles be inserted once the two liquid bodies corresponding to the detected potential collision eventually coalesce in the flow solver (if they do). Detection of this event is performed using a grouping algorithm similar to that used to bundle the collisions. For each potential collision, computational cells that lie within the region of intersection of the two bounding boxes and have a nonzero fluid volume fraction are grouped. Before a collision has occurred, the algorithm should return two groups of cells; once coalescence has occurred, the algorithm should return exactly one group of cells since the liquid-containing cells should be connected.

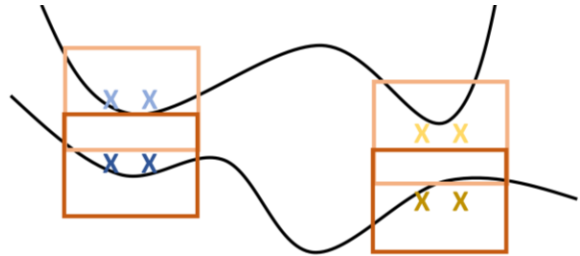

**Figure 8:** Schematic illustrating generation of bounding boxes for each half of each bundled potential collision. The bundled potential collisions shown here are identical to those in the bottom subfigure of Fig. 7.

**Demonstration of the algorithm**

In this work, the algorithm was implemented in an unstructured node-based geometric unsplit VoF code developed by the Center for Turbulence Research and Cascade Technologies (Kim *et al.*, 2014). Several test cases of increasing complexity are presented here to illustrate the performance of the algorithm.

The first test setup employed is a water drop released onto a planar water surface at atmospheric conditions. The algorithm was tested for the grid spacing $h = D/16$, where $D$ is the droplet diameter. A cubic Cartesian grid with side length $5D$ was used, and the drop was released with We = 5.7 and St = $4.0 \times 10^5$, close to the range of dimensionless parameters where microbubbles are expected to be produced. In Fig. 9, snapshots of the side view of the domain before and after a potential collision was detected are presented. Note that the positions of the mesh nodes where a potential collision was detected respect the axisymmetry of the original drop-pool system.

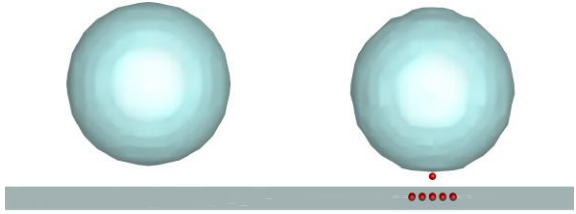

**Figure 9:** Snapshots of the computational domain before (left) and after (right) a potential collision was detected (drop on planar liquid surface). The blue surfaces represent the VoF=0.5 isosurface, and the red spheres indicate the positions of the mesh nodes where a potential collision was detected.

The next test setup employed is the head-on collision of two drops of different sizes. The We of the two drops are 7 and 10, and the St of the drops are $5.0 \times 10^5$ and $7.0 \times 10^5$. The diameter of the smaller drop is resolved with 20 cells. In Fig. 10, snapshots of the side view of the domain before and after a potential collision was detected are presented. Note, again, the symmetry of the position of the mesh nodes where a potential collision was detected and of the system itself along the line of centers of the drops, demonstrating that the collision detection algorithm respects the symmetry of the interacting features.

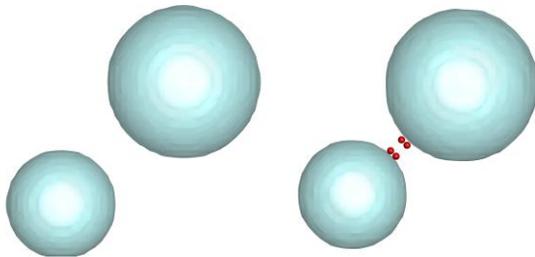

**Figure 10:** Snapshots of the computational domain before (left) and after (right) a potential collision was detected (two drops colliding head-on). See Fig. 9 for an explanation of the surfaces and features in the figure.

Another test setup that was carried out comprises randomly seeded drops of random sizes moving at random speeds in a fixed direction. 40 drops with a maximum We of about 60 and a maximum St of about $10^6$ were released in a domain with about 1.7 million cells. Snapshots of a close-up view of two potential collisions upon their detection, as well as the subsequent coalescence of the corresponding interfaces at a later time, are shown in Figs. 11 and 12.

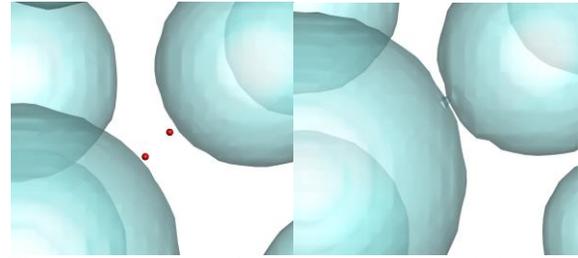

**Figure 11:** Snapshots of the computational domain just after a potential collision was detected (left) and after the corresponding interfaces eventually coalesced some time later (right) for the many-drops test case. See Fig. 9 for an explanation of the surfaces and features in the figure.

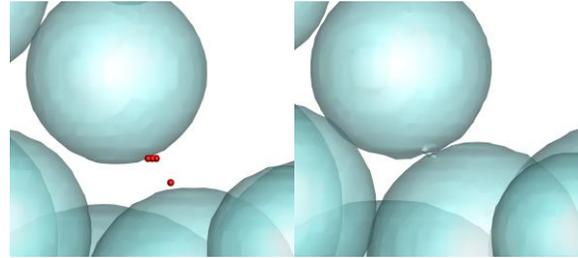

**Figure 12:** Snapshots of the computational domain just after a potential collision was detected (left) and after the corresponding interfaces eventually coalesced some time later (right) for the many-drops test case. See Fig. 9 for an explanation of the surfaces and features in the figure.

Finally, the algorithm was tested on a breaking Stokes wave (air-water) with dimensionless parameters $We = 1.6 \times 10^3$ and $Re = 1.8 \times 10^5$ using the wavelength $\lambda$ and corresponding velocity $\sqrt{g\lambda/2\pi}$ as the characteristic length and velocity scales in Eq. (3). In this formulation, Re can be defined as St [Eq. (4)] multiplied by the liquid-gas viscosity ratio [Eq. (2)], and describes the ratio of the inertial forces to the viscous forces in the system. Snapshots of the simulation before and after the wave breaking are highlighted in Fig. 13. The computational mesh consisted of about 4.2 million mesh nodes, the minimum grid resolution was $1/216$ the wavelength, and the length of the computational domain in each Cartesian axis was equal to the wavelength. The initial conditions and physical parameters used are identical to those of Wang *et al.* (2016). The first detected potential collision event is illustrated in Fig. 14.

**Computing the effective curvature**

While not formally part of the collision detection algorithm, the computation of the effective curvature of the colliding interfaces is crucial to an effective characterization of the model problem. Work is also ongoing to draw links between the detected collision events and the breakup processes investigated in the next section using the effective curvature of detected collision events. Thus, a couple of remarks on the computation of the effective curvature are in order.

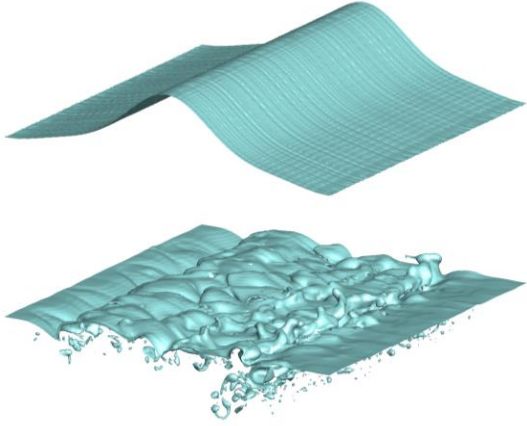

**Figure 13:** Snapshots of a bird's eye view of the VoF=0.5 isosurface for a breaking Stokes wave at two time instances: initially (top); and after about 3.6 characteristic times (bottom).

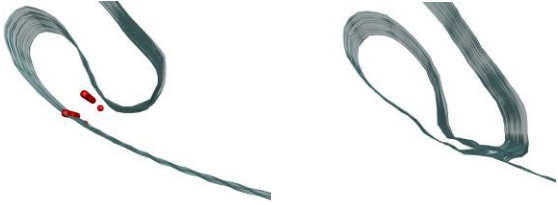

**Figure 14:** Snapshots of the computational domain just after a potential collision was detected (left) and after the corresponding interfaces eventually coalesced some time later (right) for the breaking Stokes wave test case. See Fig. 9 for an explanation of the surfaces and features in the figure.

In general, a phase interface in three-dimensional space has two principal curvatures and a velocity associated with it. Two approaching interfaces will thus have four principal curvatures and two velocities associated with them. It is straightforward to transform the two velocities into a single relative velocity, but not to transform the four curvatures into a single effective curvature. The relative velocity and effective curvature are eventually substituted into Eqs. (3) and (4) to obtain effective impact We and St for the model problem. Here, a physical transformation compatible with the physics of the flow in the trapped gas film that enables this parameter reduction is discussed.

To motivate this transformation, a closer analysis of the time evolution of the drop-pool model problem is briefly undertaken. Prior to significant deformation of the interfaces, the falling drop remains approximately spherical and the pool surface remains approximately flat. At the point of transition from the macroscale flow solver to the microscale model problem, the gas film is on the verge of becoming poorly resolved. At a resolution typical of LES ( $We_\Delta \gtrsim 1$ ), one should not expect significant deformation of the interfaces at this stage, since surface tension forces remain weak at the grid scale. Then, a Taylor expansion of the gap height near the point of closest approach can be performed, and the drop surface can be approximated as a paraboloid near this point. An approximate expression for the gap height $z$ near the point of closest approach as a function of the radial coordinate $r$ is

$$z = z_0 + \frac{r^2}{2R}, \tag{12}$$

where $z_0$ is the gap height at the point of closest approach, and $R$ is the radius of the drop. At a later stage in the approach process, an analysis of the length and velocity scales in the system permits (Smith *et al.*, 2003; Korobkin *et al.*, 2008; Hicks and Purvis, 2011; Hendrix *et al.*, 2016) the application of the lubrication approximation to the gas layer at sufficiently small interfacial separations. In particular, $z_0 \sim r^2/2R$. Under this approximation, the variation of the flow in the radial direction is small, and primarily depends on the variation of the gap height along the radial coordinate. If Eq. (12) is applied at an appropriate time to define the initial conditions for the lubrication problem, then the drop curvature used in this application of Eq. (12) directly influences the solution to the lubrication problem. Since the lubrication problem depends only on the radial variation of the gap height, it then follows that initializing the problem with a set of macroscale curvatures that results in the same effective drop curvature in Eq. (12) should yield a comparable solution.

It turns out that in the case of two arbitrarily curved interfaces, the analysis above can indeed be generalized (Derjaguin, 1934; White, 1983; Vinogradova, 1996), and the Taylor expansion about the point of closest approach can be carried out as a function of the four curvatures $R_1$, $R_2$, $R_3$ and $R_4$

$$z = z_0 + \frac{r^2}{2R_{\text{eff}}}, \tag{13}$$

where

$$\frac{1}{R_{\text{eff}}} = \sqrt{\left(\frac{1}{R_1}+\frac{1}{R_3}\right)\left(\frac{1}{R_2}+\frac{1}{R_4}\right) + \sin^2\phi \left(\frac{1}{R_1}-\frac{1}{R_2}\right)\left(\frac{1}{R_3}-\frac{1}{R_4}\right)}.$$

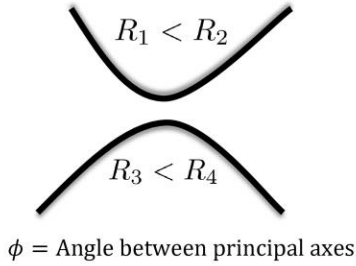

(14)

$\phi$ = Angle between principal axes

**Figure 15:** Schematic illustrating approach of two arbitrarily curved liquid-gas interfaces.

Here, $R_1 < R_2$ are the radii of curvature of the first interface, and $R_3 < R_4$ are the radii of curvature of the second interface, as illustrated in Fig. 15. The angle between the principal axes of the two interfaces at the point of closest approach is defined as $\phi$. Alternatively, this is the angle between the planes of principal curvature of the two interfaces at the point of closest approach. Eq. (14) has also been termed the Derjaguin approximation (White, 1983), and is only valid if the argument of the square root operation is positive, or when the interfaces are diverging away from and near the point of closest approach. Note that this does not preclude some of the radii of curvature from being negative, as long as $z$ is in general increasing with $r$ for small $r$. Note, also, that if all four curvatures are negative, then the interfaces are converging in the first approximation and the transformation is not desirable. From Eqs. (12) and (14), as well as the exposition above, it is evident that physical correspondence between the drop-pool model problem and the configuration of arbitrarily curved interfaces holds insofar as the gap height expressed in terms of measured quantities from the macroscale solver is a good estimate for the gap height at the point when the lubrication approximation becomes sufficiently valid (i.e., when the gap is sufficiently thin). Returning full circle to the collision detection algorithm, this again justifies the need to delay the detection of a potential collision as much as numerically possible (provided $We_\Delta \gtrsim 1$ and deformation is insignificant) for correspondence of the instant of collision detection with impact conditions.

## NUMERICAL INVESTIGATION OF SUB-HINZE SCALE BUBBLES
### Bubble size distributions and the Hinze scale

Before launching into the details of the numerical investigation of sub-Hinze scale bubbles that was performed, it is instructive to take a look at the experimentally observed properties of bubble size distributions in breaking waves, and the relation of these distributions to the Hinze scale.

In a seminal experiment on turbulent two-phase flows, Deane and Stokes (2002) performed optical measurements of bubble sizes from breaking waves in a wave flume, where the dimensionless parameters associated with the wave-packet center are $We = 1.1 \times 10^5$ and $Re = 4.4 \times 10^6$ using, again, the wavelength $\lambda$ and corresponding velocity $\sqrt{g\lambda/2\pi}$ as the characteristic length and velocity scales. About two characteristic times ($2\sqrt{\lambda/g}$) after the wave had broken, they reported an instantaneous bubble size distribution comprising two power law distributions meeting at a nondimensional length of $4 \times 10^{-4}$, which they postulated to be the Hinze scale (Hinze, 1955) of the system. A similar distribution was obtained by time-averaging the instantaneous distributions over these two characteristic times. At early times (labeled the "Jet" phase in their paper), bubbles smaller than the Hinze scale were observed to emerge first; after almost one characteristic time (labeled the "Cavity" phase in their paper), bubbles larger than the Hinze scale were then also observed following the rupture of a large cavity in the system. Isolated bubble fragmentation events were also observed and tracked. All of the observed fragmenting bubbles were larger than the Hinze scale, and none of the events generated fragmentation products significantly smaller than the Hinze scale. These observations suggest that the formation of many of the bubbles larger than the Hinze scale is governed by fragmentation due to turbulent velocity fluctuations.

Hinze (1955) postulated that in the limit of sufficiently low concentration of the dispersed phase such that random coalescence events do not occur, and for sufficiently large Reynolds numbers such that the smallest turbulent eddies in the continuous phase are much smaller than the largest entities of the dispersed phase in the system (so viscous effects are not dominant), the dominance of inertial forces due to turbulent fluctuations over capillary forces ($We_n > 1$) results in the fragmentation of large features of the dispersed phase. This cascade is terminated at what has now been termed the Hinze scale ($n = H$), where the two forces above are balanced. If the continuous phase is liquid and the dispersed phase is gaseous, as will be assumed for the rest of this discussion, then bubbles at this scale will have characteristic sizes of the order $l_n = l_H$ and will be subjected to characteristic hydrodynamic pressure fluctuations of the same order as the capillary pressure, $\rho_l u_n^2 = \rho_l u_H^2 \sim \sigma/l_H$, such that $We_H \sim 1$. Suppose, in addition, that the turbulent flow field in the mixed phase region is approximately locally isotropic. Then, one could adopt the Kolmogorov scalings typically applied to the inertial subrange of isotropic turbulence. If one assumes that

$$u_n^2 \sim (\epsilon l_n)^{2/3}, \tag{15}$$

where $\epsilon$ is the average rate of dissipation, then one could rearrange Eqs. (5), (7) and (15) to write, for $\text{We}_\text{H} \sim 1$,

$$l_\text{H} \sim (\sigma/\rho_l)^{3/5} \epsilon^{-2/5}. \tag{16}$$

Now, one could further assume that the characteristic dissipation rate is determined by the energy-containing scales, so $\epsilon \sim g^{3/2} \lambda^{1/2}/(2\pi)^{3/2}$. Note that the length scale that enters the dissipation rate scale should technically be a function of the wave height from energy considerations, but for slopes of order 0.1 to 1, this distinction does not significantly impact the derivation here. By substituting this expression for the dissipation rate into Eq. (16) and nondimensionalizing $l_\text{H}$ with $\lambda$, the nondimensional Hinze scale

$$\frac{l_\text{H}}{\lambda} \sim \left(\frac{2\pi\sigma}{\rho_l g \lambda^2}\right)^{3/5} \sim \text{We}^{-3/5} \tag{17}$$

is obtained. For the experiment by Deane and Stokes (2002), substitution of the wave-packet center parameters yields $\frac{l_\text{H}}{\lambda} \cong 9 \times 10^{-4}$ assuming a constant of proportionality of 1, in reasonable agreement with the experimentally observed break in the power-law distributions. If the critical $\text{We}_n$ for the cascade termination exceeds 1, as Hinze suggested, then the resultant prediction of $l_\text{H}/\lambda$ will be larger. It was noted earlier in this section that this analysis only holds when viscous effects are not dominant at the scales being considered. This is equivalent to the statement that the Hinze scale is larger than the Kolmogorov scale in this setting, as was remarked earlier. One could take the ratio of the Hinze scale $l_\text{H}$ to the Kolmogorov scale $l_\text{K}$ and obtain, using Eq. (17) and the scaling $l_\text{K}/\lambda \sim \text{Re}^{-3/4}$,

$$\frac{l_\text{H}}{l_\text{K}} \sim \text{We}^{-3/5} \text{Re}^{3/4}. \tag{18}$$

For the flow considered by Deane and Stokes (2002), this ratio is approximately 90, and for the breaking wave to be considered in the next section, this ratio is approximately 100, so this assumption is satisfied for these flows.

As shown by Garrett *et al.* (2000), one could extend the inertial subrange argument to the bubble size distribution directly. This extension, as well as the preceding derivation of the Hinze scale, involves assuming that a bubble interacts most intimately with an eddy of the same size, so that the bubble and eddy sizes can be interchanged in the inertial subrange scalings. Garrett *et al.* (2000) do not directly show this, so a physical justification is offered as follows. In order for a bubble to be broken by an eddy, the turnover time of the eddy $t_n \sim l_n/u_n$ has to be equal to or larger than the time involved in pinching off the bubble, which, at high Weber numbers, is of the order $t_{b,n} \sim a_n/u_n$, where $a_n$ is the characteristic size of the bubble. As a result, the turbulent eddy can only break up bubbles whose size is equal to or smaller than the eddy size, $a_n \leq l_n$. An additional condition for breakup is that the hydrodynamic pressure fluctuations generated by the eddy, $\rho_l u_n^2$, must be larger than the capillary pressure $\sigma/a_n$ in order to render a Weber number larger than 1 at that scale. From Eq. (15), this translates into the relation $\sigma/[\rho_l(\epsilon l_n)^{2/3}] \leq a_n \leq l_n$ for breakable bubbles. Note, however, that the lower bound of this range corresponds to a bubble of unity Weber number interacting with the aforementioned eddy. As a consequence, the smaller the bubble size is within this interval of breakable bubbles, the less energetically favorable the breakup process. It is therefore implied that, with high probability, an eddy of size $l_n$ will preferentially tend to break up bubbles of the same size since such bubbles render the maximum Weber number approachable during breakup. In this way, the eddy and bubble sizes can be interchangeably used in preceding and subsequent relations. Hinze (1955) arrives at the same conclusion using an alternative argument but considers neither the finite lifetime nor the finite spatial extent of an eddy. Then, assuming a constant energy cascade rate $\epsilon$ as was necessary for the establishment of an inertial subrange, as well as a constraint on the time scale of traversal through the cascade (i.e., a constant rate of air entrainment $Q$), one will find that the bubble size distribution $N(a)$ for bubbles larger than the Hinze scale depends only on $a$, $\epsilon$ and $Q$. Finally, by dimensional arguments, one could show that $N(a)$ varies as $a^{-10/3}$, a result consistent with the experimental data of Deane and Stokes (2002). However, no mechanistic explanation has been offered for the $a^{-3/2}$ scaling observed by Deane and Stokes in the same paper for sub-Hinze scale bubbles, as the authors themselves note in their discussion.

**Numerical time-averaged bubble size distributions and variation of the Hinze scale**

To numerically investigate the mechanisms of formation of bubbles of various scales in a breaking wave, one would ideally set up a canonical problem that contains the key features of a physical breaking wave with sufficient energy for realism and feasibility of comparison with experiments, but yet can be simulated with reasonable computational cost to permit ensemble averaging and long time integration. The Stokes wave, which has been simulated in 2D by Chen *et al.* (1999) and Iafrati (2009), and in 3D by

Wang *et al.* (2016) and Deike *et al.* (2016), immediately comes to mind. The Stokes wave is a periodic surface wave, allowing easy treatment of its boundary conditions. In addition, high-order Stokes waves, which involve superposition of a fundamental wave and its harmonics, can exhibit nonlinear dynamics, and will in particular break when the wave steepness is sufficiently large irrespective of the geometry of the bottom bounding surface.

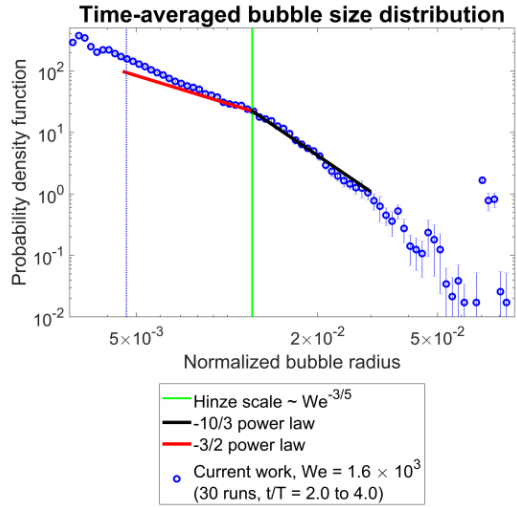

**Figure 16:** Time-averaged and ensemble-averaged bubble size distribution from a breaking Stokes wave (We = $1.6 \times 10^3$ and Re = $1.8 \times 10^5$). The blue, dotted vertical line denotes the grid resolution, while the green, solid vertical line denotes the estimated Hinze scale based on Eq. (17) with a unity proportionality constant. The black (steeper) sloped line denotes a -10/3 power law, while the red (gentler) sloped line denotes a -3/2 power law. The error bars denote the 95% confidence interval over the ensemble for each bubble size bin. The estimated Hinze scale appears to coincide with the break between the power law fits. The data points above the Hinze scale appear to follow the -10/3 power law more closely, suggesting the presence of a turbulent fragmentation cascade in this region. The data points below the Hinze scale appear to follow the -3/2 power law more closely, suggesting the resolution of several sub-Hinze scale bubbles, although there is some observable deviation from the -3/2 power law.

In this work, a 27-cm-long water wave at atmospheric conditions with a wave steepness of 0.55, where the wave steepness is the product of the wave amplitude and the fundamental wavenumber, was simulated. This corresponds to the dimensionless parameters We = $1.6 \times 10^3$ and Re = $1.8 \times 10^5$, which were also used by Wang *et al.* (2016) in their simulations, and are identical to the final test case of the collision detection algorithm discussed above. The corresponding initial velocity field is documented in Iafrati (2009). Periodic boundary conditions were employed in the streamwise and spanwise directions, and slip boundary conditions were employed in the wave-normal direction. The reader is referred to the aforementioned discussion for more details of this simulation.

In order to construct the bubble size distributions that follow, a grouping algorithm (flood-fill) similar to that used by Wang *et al.* (2016) and Deike *et al.* (2016) was used to identify contiguous groups of air-containing cells. It may be shown that this algorithm identifies spurious bubbles since the turbulent breakup process generates closely-spaced air pockets with small air volume fractions that are grouped together by this algorithm. In this study, air-containing cells are grouped together only when at least one of the cells contains a significant amount of air, in order to circumvent the identification of these spurious bubbles, as well as poorly-resolved bubbles.

The time-averaged bubble size distribution over the active wave breaking period, taken approximately to be the time interval between the rupture of the first air cavity that spans the entire domain and the estimated point in time at which the population of the largest bubbles starts to decrease, is plotted against the nondimensional radii of the bubbles in Fig. 16. For this plot and for all subsequent statistics in this work, an ensemble average over 30 runs was performed with small perturbations to the initial interface. The plot suggests that the resolution of the performed simulations is just sufficient to resolve the Hinze scale, and some sub-Hinze scale bubbles are resolved over a limited size range. The super-Hinze scale bubbles are well-resolved and appear to converge more closely to the -10/3 power law than the -3/2 power law, suggesting the presence of a cascading breakup process established by turbulent fluctuations.

Since no mechanistic explanation for the formation of sub-Hinze scale bubbles has been offered thus far, increased resolution of these bubbles in a numerical simulation of a turbulent breaking wave, followed by tracking and probing of these bubbles, may shed more light on this phenomenon. Increased resolution of sub-Hinze scale bubbles entails one of the two following actions (or a combination of them): reduction of the minimum grid size keeping the dimensionless parameters constant so that it becomes much smaller than the estimated Hinze scale, or manipulation of the dimensionless parameters to increase the estimated Hinze scale keeping the grid size constant. In this work, the latter option is attempted in order to keep computational costs manageable. As shown in the previous subsection, decreasing the We of the wave will result in an increase in the Hinze scale. A second set of simulations with We = $3.5 \times 10^2$ was performed keeping all other parameters constant. This increases the estimated Hinze scale by about 2.5 times. The time-averaged and ensemble-averaged bubble size distribution is plotted in Fig. 17. The bubbles just larger than the minimum grid size appear to follow the -3/2 power law more closely than the -10/3 power law, suggesting the presence of resolved sub-Hinze scale bubbles in the system. Note the extended size range of resolved sub-Hinze scale bubbles in Fig. 17 relative to the corresponding size range in Fig. 16.

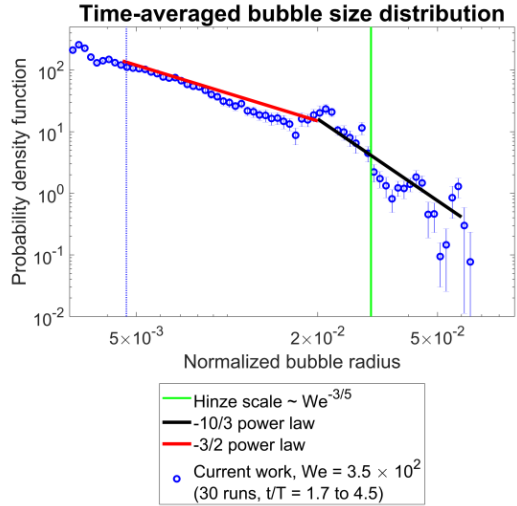

**Figure 17:** Time-averaged and ensemble-averaged bubble size distribution from a breaking Stokes wave (We = $3.5 \times 10^2$ and Re = $1.8 \times 10^5$). See Fig. 16 for an explanation of the various lines in the plot. The data points just right of the blue dotted line appear to follow the -3/2 power law more closely, suggesting the presence of sub-Hinze scale bubbles. However, as in Fig. 16, there is some observable deviation from the -3/2 power law in the sub-Hinze scale region. The size range of sub-Hinze scale bubbles appears to be more substantial relative to the corresponding size range observed in Fig. 16.

**Instantaneous bubble size distributions and comparison with flow structures**

In order to gain more insight into the time evolution of the bubble size distribution and to compare this evolution with the qualitative observations of Deane and Stokes (2002), snapshots of the ensemble-averaged time-instantaneous bubble size distribution in the early, middle and late stages of the active wave breaking period, together with corresponding snapshots of the interface from different angles, are shown in the subsequent figures. Since the flow under consideration is statistically unsteady, ensemble averaging is a more appropriate way of obtaining statistics than time averaging. In Figs. 18, 19, 20 and 21, snapshots for the baseline case (We = $1.6 \times 10^3$) are provided at nondimensional times $t^*$ of about 2.5, 3.0, 3.5 and 4.0 respectively. In Figs. 22, 23, 24 and 25, snapshots for the case with the larger Hinze scale (We = $3.5 \times 10^2$) are provided at the same nondimensional times respectively.

For the baseline case, it is evident from Fig. 18 that the smaller bubbles in the system are produced first. Many of these bubbles with sizes just above the grid resolution appear to follow the -3/2 power law more closely than the -10/3 power law at $t^* = 2.5$ as may be observed in the top panel of Fig. 18, and are likely to have been formed from the thin S-shaped film stretching out and fragmenting under the wave interface on the right of the bottom panel of Fig. 18. The S-shaped film wraps around the cylindrical air cavity on the left of the bottom panel of Fig. 18 in a clockwise fashion, resembling the phenomenology observed by Deane and Stokes (2002). At $t^* = 3.0$, the cylindrical air cavity, now in the middle of the bottom panel of Fig. 19, begins to deform under the action of turbulent eddies, and the -10/3 power law appears to instantaneously extend into the sub-Hinze scale region, as seen in the top panel of Fig. 19. At $t^* = 3.5$, the top panel of Fig. 20 suggests that more super-Hinze scale bubbles are emerging in the system, extending the range of the -10/3 power law towards larger bubble sizes. This appears to coincide with the fragmentation of the aforementioned originally-cylindrical air pocket. At $t^* = 4.0$, the sub-Hinze scale bubbles appear to instantaneously follow a -3/2 power law closely, while the super-Hinze scale bubbles appear to instantaneously follow a -10/3 power law closely. This seems to coincide with near-complete rupture of the originally-cylindrical air pocket. Note that while there is observable deviation from the -3/2 power law in the sub-Hinze scale region in the time-averaged distribution in Fig. 16, the bubble size distribution instantaneously coincides with the -3/2 power law in the sub-Hinze scale region in the top panel of Fig. 21.

For the case with the larger Hinze scale, the near-complete rupture of the cylindrical air pockets formed from the overturning wave at $t^* = 3.0$ also appears to coincide with the sub-Hinze scale bubbles instantaneously following a -3/2 power law closely and the super-Hinze scale bubbles instantaneously following a -10/3 power law closely, as seen in the top panel of Fig. 23. Note, again, that there is less deviation from the -3/2 power law in the instantaneous ensemble-averaged distribution in Fig. 23 compared to the time-averaged and ensemble-averaged distribution in Fig. 17. Because surface tension is more dominant in this test case, the wave crest impacts the rest of the water surface less forcefully as it breaks, and fewer bubbles seem to be produced. As such, the statistics for this case appear to be less converged than those for the baseline case with the same ensemble size, as is evidenced from the lengths of the 95% confidence intervals in the top panels of Figs. 22 to 25. In addition, a larger proportion of the bubbles is closer to the surface on average, as one may tell with a cursory comparison of Figs. 20 and 24. By $t^* = 4.0$, many of the large bubbles have already risen to the surface due to buoyancy, resulting in the dip in number of large bubbles evident in the top panel of Fig. 25. Visual inspection of Figs. 22 to 25, as well as Fig. 18 to some extent, suggests that capillary mechanisms like the Plateau-Rayleigh instability, thin film retraction and other pinch-off phenomena appear to be dominant in the formation of sub-Hinze scale bubbles, and more detailed analysis of the formation and evolution of individual bubbles is ongoing to provide more insight on these mechanisms.

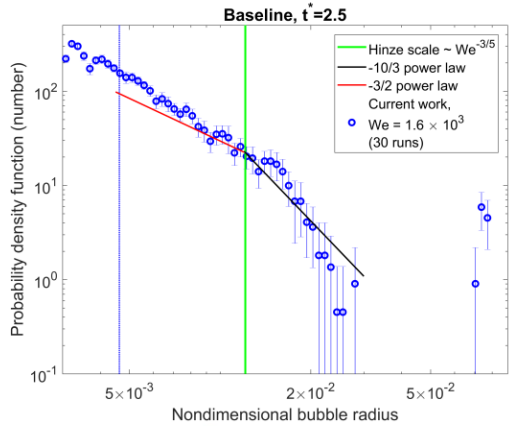
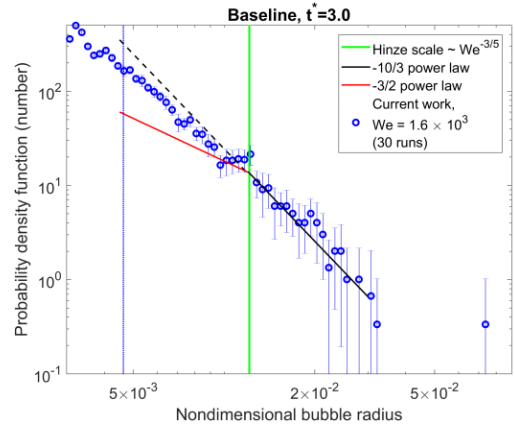
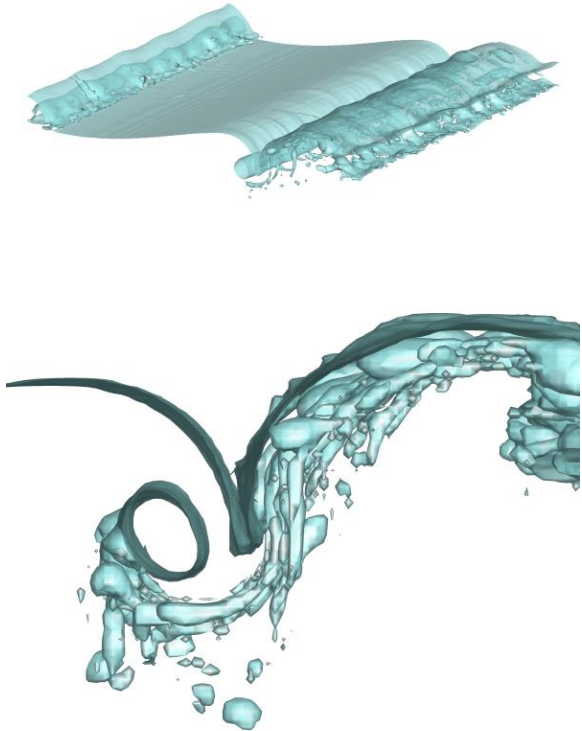
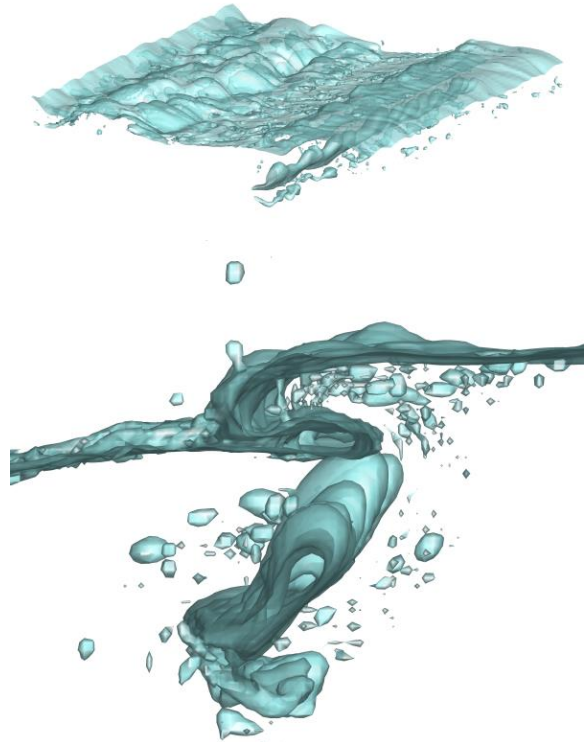

**Figure 18:** Snapshots of the ensemble-averaged instantaneous bubble size distribution (top panel), a bird's eye view of the wave interface from a single run (center panel) and a close-up side view of the same interface (bottom panel) for the baseline case at the nondimensional time $t^* = 2.5$.

**Figure 19:** Snapshots for the baseline case at $t^* = 3.0$. See Fig. 18 for a description of the panels.

## CONCLUSIONS AND FUTURE WORK

The resolution of microbubbles near air-water interfaces is expensive. Hence, in this work, it is proposed that an SGS model is necessary to accurately predict microbubbles. This work offers details on a proposed SGS model and takes an in-depth look at two components of the SGS model: the collision detection algorithm intended to activate the SGS model, and a numerical investigation of breaking waves intended to offer insights into justification and potential refinement of the selected model problem.

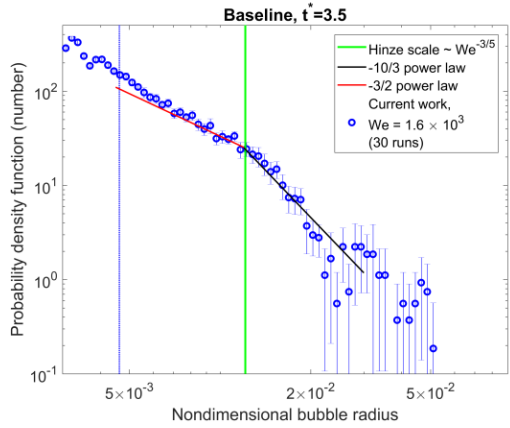
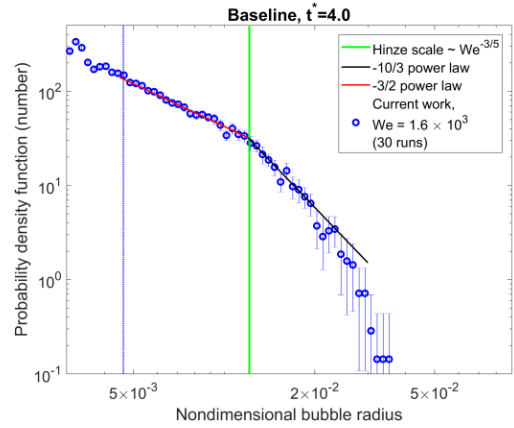
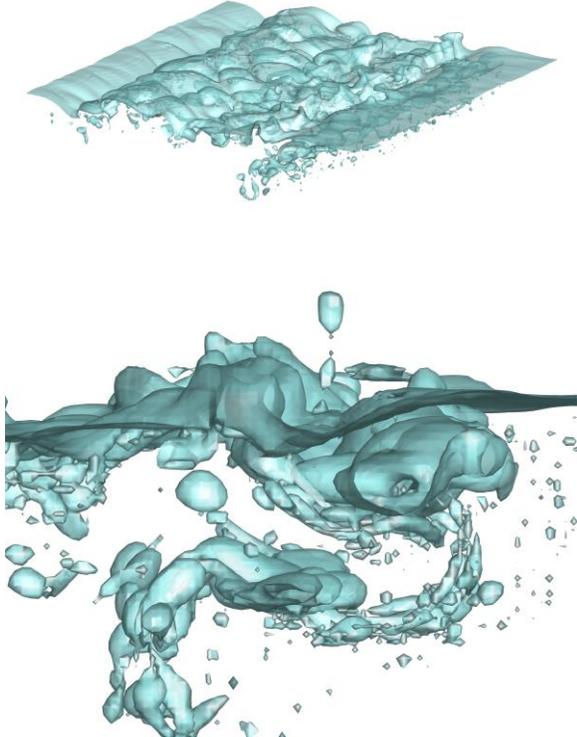
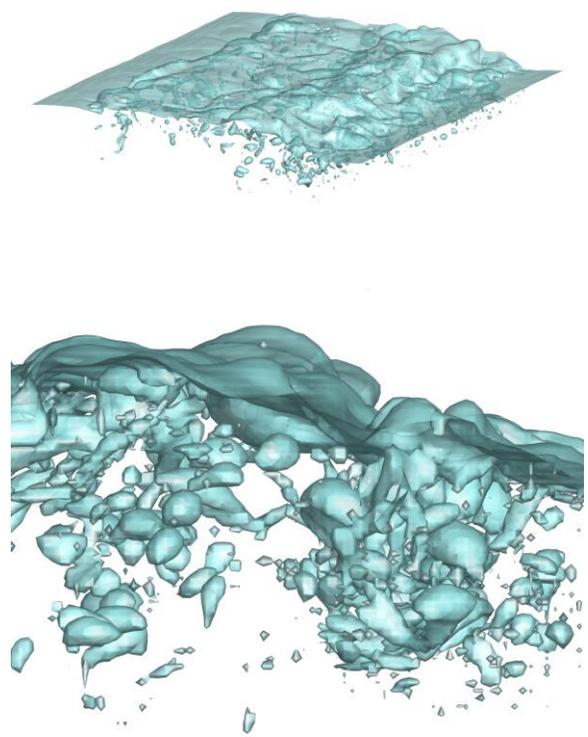

**Figure 20:** Snapshots for the baseline case at $t^* = 3.5$. See Fig. 18 for a description of the panels.

**Figure 21:** Snapshots for the baseline case at $t^* = 4.0$. See Fig. 18 for a description of the panels.

    A collision detection algorithm compatible with various interface-capturing methods and numerical schemes was developed to identify regions with a high probability of microbubble formation. Since quantities derived from the macroscale flow solver are used to compute the input parameters for the

microscale model problem, care has to be taken to ensure physical correspondence of these quantities and parameters. Triggering the transition to the model problem too early increases the temporal separation between the time of impact assumed in the macroscale flow solver and the true time of impact, but triggering the transition too late degrades the quality of the estimates of the relevant quantities from the macroscale flow solver due to numerical errors near poorly-resolved features. An appropriate transition time motivated by the order of accuracy of the numerical solver is proposed, and heuristics designed to efficiently obtain a minimal set of new and unique potential collisions every computational time step are adopted in the detection algorithm. The algorithm was implemented in a geometric VoF solver, and its performance was demonstrated for various test cases of differing physical complexity, including a breaking Stokes wave. Work is ongoing to minimize the cost of the collision detection algorithm for usage in scalable codes for the simulation of large systems without excessive compromise on the accuracy of the various heuristics. Analysis of the potential collisions yielded in the breaking Stokes wave may also offer more insights into the breakup processes occurring in the wave as it breaks and generates features of a wide range of scales.

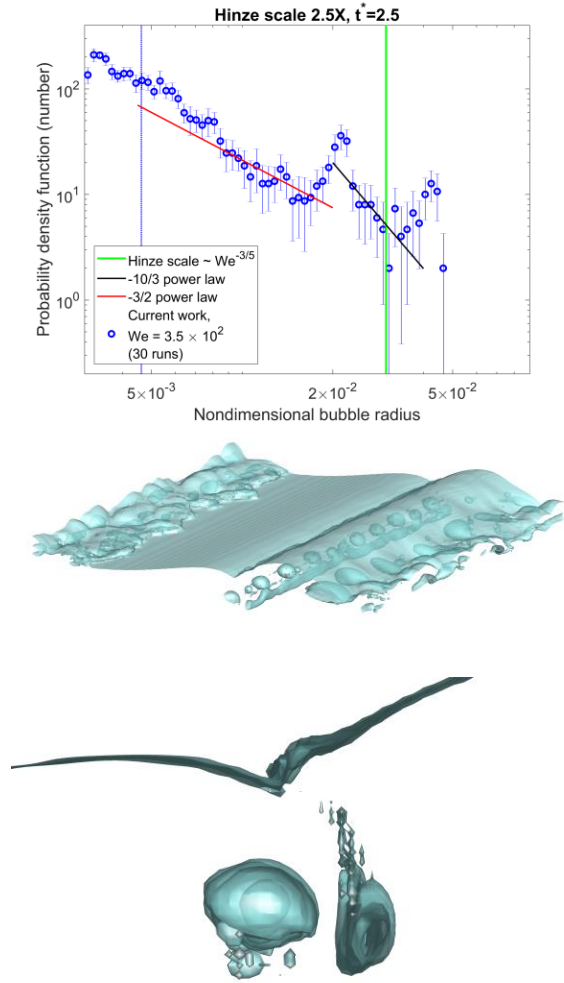

**Figure 22:** Snapshots for the larger Hinze scale case at $t^* = 2.5$. See Fig. 18 for a description of the panels.

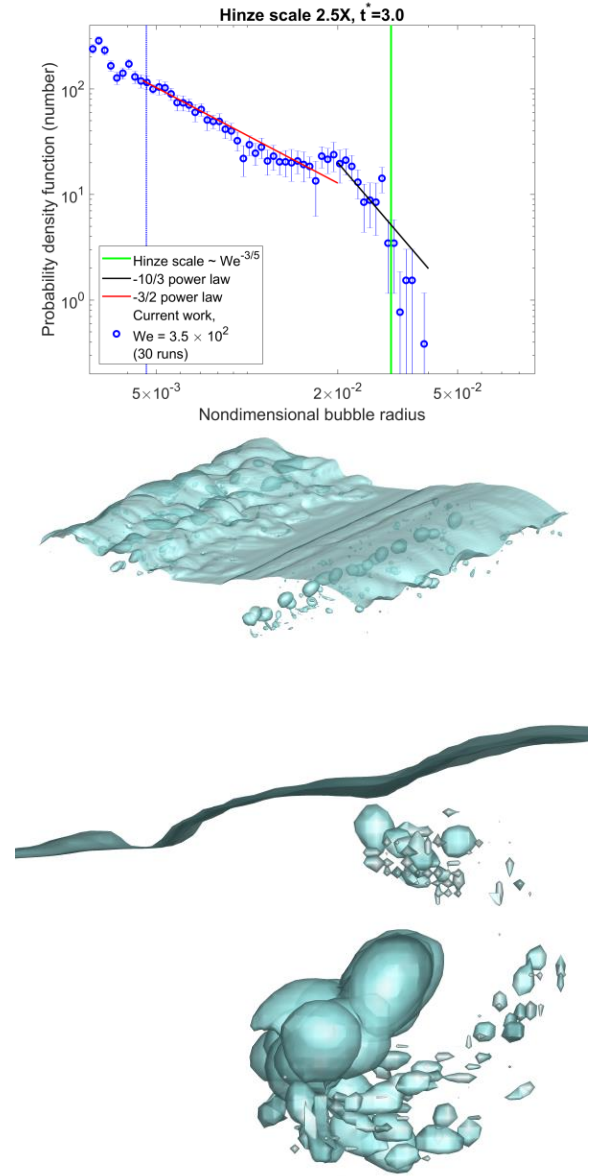

**Figure 23:** Snapshots for the larger Hinze scale case at $t^* = 3.0$. See Fig. 18 for a description of the panels.

Experiments of breaking waves have yielded time-averaged and instantaneous bubble size distributions that are comprised of two power-law distributions. It has been postulated that the transition between the two distributions occurs at the Hinze scale, where the hydrodynamic pressure fluctuations due to turbulence are of the same order as the capillary pressure at that scale. Above the Hinze scale, the bubble size distribution resembles a $-10/3$ power law, and below the Hinze scale, the distribution resembles a $-3/2$ power law. Assuming that air is entrained by the breaking wave at a constant rate and is then subject to the action of turbulent eddies that are approximately locally isotropic, it can be shown from dimensional arguments that a $-10/3$ power law distribution for the bubble sizes is expected when air entrainment is active. No similar mechanistic argument, however, has been offered for the $-3/2$ power law, which incidentally describes the microbubbles that have not all been resolved even with today's state-of-the-art simulations. In order to capture capillary effects that may be crucial in the formation of these microbubbles, it is expected that a true DNS requires a grid resolution far smaller than the Hinze scale for highly energetic waves, independent of the need to resolve the Kolmogorov scale. This necessitates the development of an SGS model for sub-Hinze scale bubbles in the simulation of highly energetic flows. In this study, the ratio of the characteristic length scales is adjusted for a numerical simulation of the breaking Stokes wave configuration such that the wave is slightly less energetic than in previously carried out experiments and numerical simulations, but also such that the grid resolves more sub-Hinze scale bubbles than in the baseline case. This is done to provide the opportunity to probe and analyze the formation of more sub-Hinze scale bubbles amid turbulent breakup processes, in the hope of obtaining a better understanding of the origin of these sub-Hinze scale bubbles. A refined bubble identification algorithm was developed in tandem to ensure that only physical bubbles are considered in the interpretation of the bubble size distributions that result from the simulations. Preliminary results from simulation of this configuration suggest that a larger number of bubbles larger than the grid resolution do indeed obey a $-3/2$ power law more closely than a $-10/3$ power law as compared to the baseline case, and thus that potentially more sub-Hinze scale bubbles are being resolved in this configuration. A closer analysis of the instantaneous ensemble-averaged bubble size distribution, as well as visual inspection of the phase interfaces in the system, reveals, in accordance with experimental observations, that the distribution of the bubble sizes, especially of the larger bubbles, changes significantly over a time interval of one to two characteristic times, and thus that the time-averaged bubble size distribution may not offer the full picture on the dynamics of bubble formation. They indicate that the formation of bubbles much larger than the Hinze scale mostly occurs after the formation of bubbles smaller than the Hinze scale. Also, they suggest that the presence of a $-3/2$ power law is associated with capillary mechanisms, including the Plateau-Rayleigh instability, thin film retraction and other pinch-off phenomena. Work is ongoing to investigate this hypothesis more closely, as well as to investigate the role of coalescence in the development of these distributions.

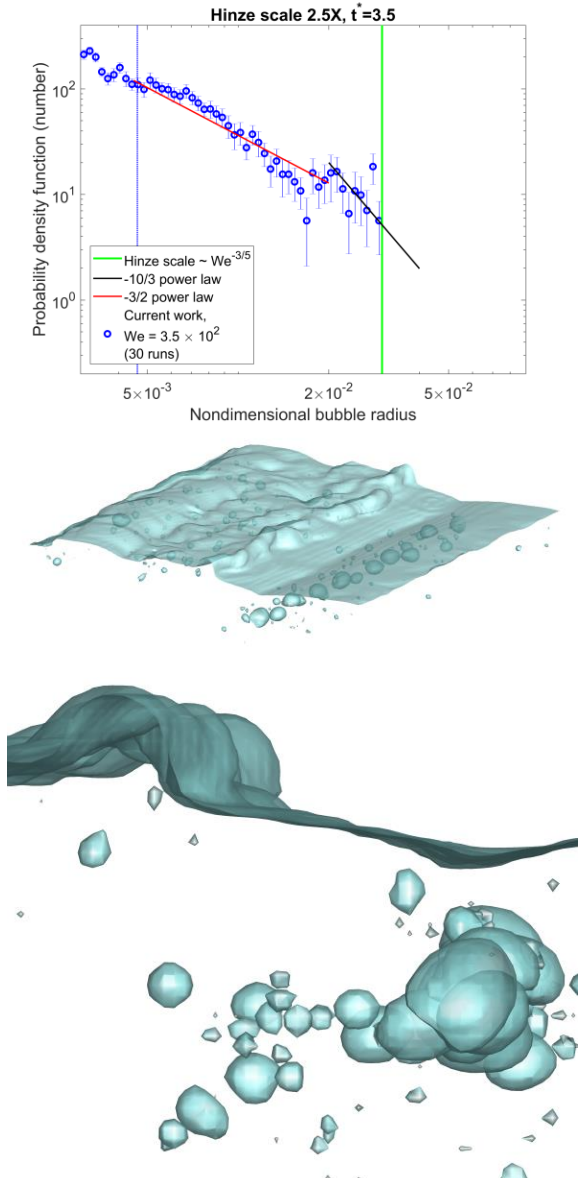

**Figure 24:** Snapshots for the larger Hinze scale case at $t^* = 3.5$. See Fig. 18 for a description of the panels.

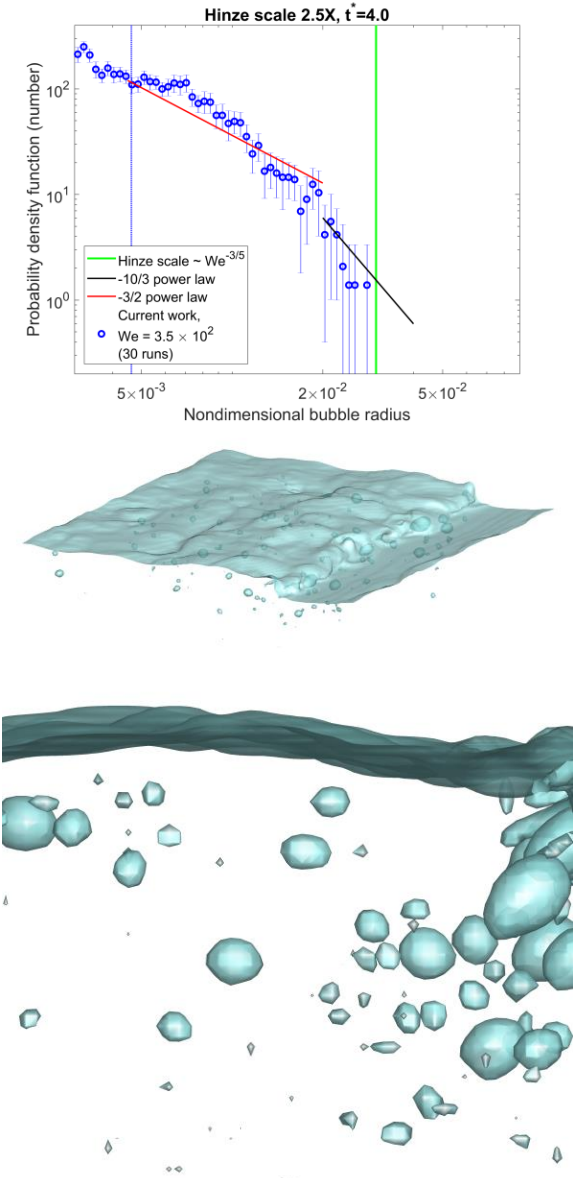

**Figure 25:** Snapshots for the larger Hinze scale case at $t^* = 4.0$. See Fig. 18 for a description of the panels.

## ACKNOWLEDGEMENTS

This work is supported by the U. S. Office of Naval Research (ONR), Grant #N00014-15-1-2726. The program manager is Dr. Thomas Fu. W. H. R. Chan is also supported by the Agency of Science, Technology and Research (A*STAR), Singapore. The authors are grateful to Shahab Mirjalili and Prof. Ali Mani for discussions on the SGS model, and to Dr. Milad Mortazavi for input on the collision detection algorithm, as well as Dr. Michael Dodd for fruitful discussions on the analysis of the bubble size distributions. The authors also thank Dr. Christopher Ivey, Dr. Dokyun Kim, Dr. Lluís Jofre and Dr. Sanjeeb Bose for technical assistance in the VoF simulations.

## REFERENCES

Baldessari, F., Homsy, G. M. and Leal, L. G., "Linear stability of a draining film squeezed between two approaching droplets," Journal of Colloid and Interface Science, Vol. 307, 2007, pp. 188–202.

Blanchard, D. C. and Woodcock, A. H., "Bubble formation and modification in the sea and its meteorological significance," Tellus, Vol. 9, 1957, pp. 145–158.

Bouwhuis, W., van der Veen, R. C. A., Tran, T., Keij, D. L., Winkels, K. G., Peters, I. R., van der Meer, D., Sun, C., Snoeijer, J. H. and Lohse, D., "Maximal air bubble entrainment at liquid-drop impact," Physical Review Letters, Vol. 109, 2012, 264501.

Carroll, K. and Mesler, R., "Part II: bubble entrainment by drop-formed vortex rings," AIChE Journal, Vol. 27, 1981, pp. 853–856.

Chan, W. H. R., Urzay, J., Mani, A. and Moin, P., "On the development of a subgrid-scale model for the generation of micro-bubbles from impacting liquid surfaces," Center for Turbulence Research Annual Research Briefs, Stanford University, 2016, pp. 149–162.

Chan, W. H. R., Urzay, J. and Moin, P., "Development of a collision detection algorithm for turbulent two-phase flows," Center for Turbulence Research Annual Research Briefs, Stanford University, 2017, pp. 103–116.

Chen, G., Kharif, C., Zaleski, S. and Li, J., "Two-dimensional Navier-Stokes simulation of breaking waves," Physics of Fluids, Vol. 11, 1999, pp. 121–133.

Culick, F. E. C., "Comments on a ruptured soap film," Journal of Applied Physics, Vol. 31, 1960, pp. 1128–1129.

Deane, G. B. and Stokes, M. D., "Scale dependence of bubble creation mechanisms in breaking waves," Nature, Vol. 418, 2002, pp. 839–844.

Deike, L., Melville, W. K. and Popinet, S., "Air entrainment and bubble statistics in breaking waves," Journal of Fluid Mechanics, Vol. 801, 2016, pp. 91–129.

Derjaguin, B. V., "Untersuchngen über die Reibung und Adhäsion, IV," Kolloid-Zeitschrift, Vol. 69, 1934, pp 155–164.

Duchemin, L. and Josserand, C., "Curvature singularity and film-skating during drop impact," Physics of Fluids, Vol. 23, 2011, 091701.

Esmailizadeh, L. and Mesler, R., "Bubble entrainment with drops," Journal of Colloid and Interface Science, Vol. 110, 1986, pp. 561–574.

Garrett, C., Li, M. and Farmer, D., "The connection between bubble size spectra and energy dissipation rates in the upper ocean," Journal of Physical Oceanography, Vol. 30, 2000, pp. 2163–2171.

Ham, F., Mattsson, K. and Iaccarino, G., "Accurate and stable finite volume operators for unstructured flow solvers," Center for Turbulence Research Annual Research Briefs, Stanford University, 2006, pp. 243–261.

Hendrix, M. H. W., Bouwhuis, W., van der Meer, D., Lohse, D. and Snoeijer, J. H., "Universal mechanism for air entrainment during liquid impact," Journal of Fluid Mechanics, Vol. 789, 2016, pp. 708–725.

Hicks, P. D. and Purvis, R., "Air cushioning in droplet impacts with liquid layers and other droplets," Physics of Fluids, Vol. 23, 2011, 062104.

Hinze, J. O., "Fundamentals of the hydrodynamic mechanism of splitting in dispersion processes," AIChE Journal, Vol. 1, 1955, pp. 289–295.

Iafrati, A., "Numerical study of the effects of the breaking intensity on wave breaking flows," Journal of Fluid Mechanics, Vol. 622, 2009, pp. 371–411.

Kaur, S. and Leal, L. G., "Three-dimensional stability of a thin film between two approaching drops," Physics of Fluids, Vol. 21, 2009, 072101.

Kim, D., Ham, F., Bose, S., Le, H., Herrmann, M., Li, X., Soteriou, M. C. and Kim, W., "High-fidelity simulation of atomization in a gas turbine injector high shear nozzle," ILASS Americas: Proceedings of the 26th Annual Conference, 2014, Portland, Oregon.

Korobkin, A. A., Ellis, A. S. and Smith, F. T., "Trapping of air in impact between a body and shallow water," Journal of Fluid Mechanics, Vol. 611, 2008, pp. 365–394.

Le Chenadec, V., Mirjalili, S., Mortazavi, M. and Mani, A., "Feature identification algorithms for sharp-interfaces flows," Center for Turbulence Research Annual Research Briefs, Stanford University, 2014, pp. 59–67.

Mandre, S. and Brenner, M. P., "The mechanism of a splash on a dry solid surface," Journal of Fluid Mechanics, Vol. 690, 2012, pp. 148–172.

Mani, M., Mandre, S. and Brenner, M. P., "Events before droplet splashing on a solid surface," Journal of Fluid Mechanics, Vol. 647, 2010, pp. 163–185.

Melville, W. K., "The role of surface-wave breaking in air-sea interaction," Annual Review of Fluid Mechanics, Vol. 28, 1996, pp. 279–321.

Mills, B. H., Saylor, J. R. and Testik, F. Y., "An experimental study of Mesler entrainment on a surfactant-covered interface: the effect of drop shape and Weber number," AIChE Journal, Vol. 58, 2012, pp. 46–58.

Mirjalili, S. and Mani, A., "High fidelity simulations of micro-bubble shedding from retracting thin gas films in the context of liquid-liquid impact," Proceedings of the 32nd Symposium on Naval Hydrodynamics, 2018, Hamburg, Germany.

Mortazavi, M., Le Chenadec, V., Moin, P. and Mani, A., "Direct numerical simulation of a turbulent hydraulic jump: turbulence statistics and air entrainment," Journal of Fluid Mechanics, Vol. 797, 2016, pp. 60–94.

Prosperetti, A. and Oğuz, H. N., "The impact of drops on liquid surfaces and the underwater noise of rain," Annual Review of Fluid Mechanics, Vol. 25, 1993, pp. 577–602.

Pumphrey, H. C. and Elmore, P. A., "The entrainment of bubbles by drop impacts," Journal of Fluid Mechanics, Vol. 220, 1990, pp. 539–567.

Reed, A. M. and Milgram, J. H., "Ship wakes and their radar images," Annual Review of Fluid Mechanics, Vol. 34, 2002, pp. 469–502.

Salerno, E., Levoni, P. and Barozzi, G. S., "Air entrainment in the primary impact of single drops on a free liquid surface," Journal of Physics: Conference Series, Vol. 655, 2015, 012036.

Sigler, J. and Mesler, R., "The behavior of the gas film formed upon drop impact with a liquid surface," Journal of Colloid and Interface Science, Vol. 134, 990, pp. 459–474.

Smith, F. T., Li, L. and Wu, G. X., "Air cushioning with a lubrication/inviscid balance," Journal of Fluid Mechanics, Vol. 482, 2003, pp. 291–318.

Stanic, S., Caruthers, J. W., Goodman, R. R., Kennedy, E. and Brown, R. A., "Attenuation measurements across surface-ship wakes and computed bubble distributions and void fractions," IEEE Journal of Oceanic Engineering, Vol. 34, 2009, pp. 83–92.

Taylor, G. I., "The dynamics of thin sheets of fluid. III. Disintegration of fluid sheets," Proceedings of the Royal Society of London A, Vol. 253, 1959, pp. 313–321.

Thoraval, M.-J., Takehara, K., Etoh, T. G. and Thoroddsen, S. T., "Drop impact entrapment of bubble rings," Journal of Fluid Mechanics, Vol. 724, 2013, pp. 234–258.

Thoroddsen, S. T., Etoh, T. G. and Takehara, K., "Air entrainment under an impacting drop," Journal of Fluid Mechanics, Vol. 478, 2003, pp. 125–134.

Thoroddsen, S. T., Etoh, T. G., Takehara, K., Ootsuka, N. and Hatsuki, Y., "The air bubble trapped under a drop impacting on a solid surface," Journal of Fluid Mechanics, Vol. 545, 2005, pp. 203–212.

Trevorrow, M. V., Vagle, S. and Farmer, D. M., "Acoustical measurements of microbubbles within ship wakes," The Journal of the Acoustical Society of America, Vol. 95, 1994, pp. 1922–1930.

Vinogradova, O. I., "Hydrodynamic interaction of curved bodies allowing slip on their surfaces," Langmuir, Vol. 12, 1996, pp. 5963–5968.

Wang, A.-B., Kuan, C.-C. and Tsai, P.-H., "Do we understand the bubble formation by a single drop impacting upon liquid surface?" Physics of Fluids, Vol. 25, 2013, 101702.

Wang, Z., Yang, J. and Stern, F., "High-fidelity simulations of bubble, droplet and spray formation in breaking waves," Journal of Fluid Mechanics, Vol. 792, 2016, pp. 307–327.

White, L. R., "On the Deryaguin approximation for the interaction of macrobodies," Journal of Colloid and Interface Science, Vol. 95, 1983, pp. 286–288.

Zhang, X., Lewis, M., Bissett, W. P., Johnson, B. and Kohler, D., "Optical influence of ship wakes," Applied Optics, Vol. 43, 2004, pp. 3122–3132.

Zhao, H., Brunsvold, A. and Munkejord, S. T., "Transition between coalescence and bouncing of droplets on a deep liquid pool," International Journal of Multiphase Flow, Vol. 37, 2011, pp. 1109–1119.

## DISCUSSION I

(Pablo Carrica, University of Iowa)

This excellent paper presents the modeling for microbubble entrainment generated by colliding water interfaces (Mesler entrainment). The work is significant in that seeks to model and understand how microbubbles, smaller than the Hinze scale, are entrained. These bubbles are present in ship wakes, and are very important for acoustic signatures.

Discussion items:

1) As the bubbles are injected, are these small bubbles interacting with the flow in later development? It is known that these bubbles may coalesce with each other or bigger bubbles that can be resolved by the grid. Please comment on it.

2) Assuming the model based on the retracting film is reasonable, the injected bubble size may have large uncertainty, because the gas film which is highly dependent on the grid resolution and the reconstruction of the surface. The paper states that more accurate methods are under development, but it would be interesting to know how sensitive the distribution of small bubbles is.

3) The paper shows the prediction of potential collision with surface well resolved. As shown in Fig. 19, there may be many droplets for which the surface may not be well resolved, possible only by 1 or 2 grid points. Has the collision detection algorithm been tested for such situations?

4) In Fig. 17, the authors show the bubble size distribution for large Hinze scale. It can be seen that the "resolved" region for -3/2 power law is small, approximately 4 times the grid size. In Figs. 19, 20 and 21, we do see many "bubbles" with very poor resolution. How do the authors define a bubble in this case?

5) Coalescence is a notoriously difficult process to simulate with DNS, due to the small scale of the film between the bubbles and the short time for rupture. Have the authors considered these constraints, or are the authors planning a different approach when proposing to investigate the role of coalescence at the end of the conclusions?

6) Please comment also on possible limitations of the approach; nobody better than the authors to know the intricacies of the methodology and where things could be improved. For instance, flotsam and jetsam artifacts frequently present in VOF approaches can produce unphysical small bubbles with usually negligible void fraction but important consequences to size distributions.

## AUTHORS' REPLY

Thank you for your comments and questions, each of which is addressed below.

1) Breakup models (e.g., Apte *et al.*, 2003) are crucial for Lagrangian particles with large Weber numbers. Coalescence models will be important when the volume fraction of these particles is large. These models have not yet been implemented in our work.

2) We expect the macroscale grid to resolve the impacting interfaces for high-aspect-ratio films so there should not be a large uncertainty associated with the corresponding dimensionless parameters. The dimensionless parameters are used as inputs to a model sub-problem, which when solved accurately (e.g., Mirjalili and Mani, 2018) will reduce the uncertainty of the injected sizes. Low-aspect-ratio films may involve subgrid corrugations that need to be modeled. Statements clarifying this distinction have been added to the manuscript.

3) These droplets should be transferred to the Lagrangian phase and should not be subject to the Eulerian collision detection algorithm.

4) We have added clarifications in the manuscript to address the role of the numerical bubbles in our bubble identification algorithm (modified flood-fill). These numerical bubbles are not included in the computation of our distributions.

5) We are only seeking to investigate transfer fluxes, which involve the sizes of the participating bubbles (and potentially the final bubble) and are directly manifested in the time-evolving bubble size distribution. We do not intend to resolve the dynamics of coalescence (e.g., coalescence efficiency). Spurious coalescence events could be identified by grid convergence of these transfer fluxes or predictions from higher-fidelity simulations and experiments, and removed after identification of these events via the collision detection algorithm.

6) We have added clarifications in the manuscript to address the role of the numerical bubbles in our bubble identification algorithm (modified flood-fill). These numerical bubbles are not included in the computation of our distributions. A proper subgrid-scale model should transfer these bubbles to the Lagrangian phase to be treated separately.

## DISCUSSION II

(Frederic Gibou, University of California, Santa Barbara)

This paper describes a physics-based impact and breakup model for the generation of microbubbles.

The authors have introduced an excellent approach and provided a thorough discussion of their model and computational treatments.

The author focus on a model problem of the impact of a spherical bubble falling on a deep liquid pool, which is justified by the literature. Suggestions for the discussion:

1) The size distribution is based on the breakup dynamics of a retracting gas film with a finite boundary. Would the presence of nearby impacts influence the distribution of generated microbubbles? If this is so, what modifications would the SGS model need?

2) Would complementing this analysis with a Deep Learning Techniques be of interest?

**AUTHORS' REPLY**

Thank you for your comments and questions, each of which is addressed below.

1) The current SGS model assumes collisions are well-separated in space and time. Additional constraints on the SGS model will be required for nearly-simultaneous and closely-spaced collisions. We have added a statement in the manuscript to clarify this.

2) The fidelity of the lookup table may be improved by interpolation or deep learning methods. We have added a statement in the manuscript to clarify this.

**DISCUSSION III**

(James Duncan, University of Maryland)

When I examined high-speed movies of bubble for motion in breaking flows and other turbulent flows, my eye is drawn to the splitting of larger bubbles. In addition to the two larger bubbles created by splitting, I typically see some very small bubbles created. Though I imagine that the collision processes that you mention making an important contribution to small bubble creation, I think those small bubbles created by large bubble splitting will be significant as well. Do you consider both of these small bubble generation processes to be physically similar?

**AUTHORS' REPLY**

Thank you for your question. We are still actively investigating these phenomena, but we currently believe that these small bubble generation processes are related. The inertial deformation of the neck generated during large bubble splitting may result in satellite bubble formation, in a similar (but not identical) fashion to how the viscous deformation of the film trapped between two impacting liquid surfaces eventually causes the rupture and retraction of the film (e.g., Mani *et al.*, 2010). The former phenomenon was investigated by Gordillo and Fontelos (2007) and Gordillo (2008), and has been shown to generate satellite bubbles whose radius is about $10^{-4}$ to $10^{-5}$ times the radius of the parent bubble (for air-water systems). We believe a collision detection algorithm may be able to identify both types of events for appropriate treatment.

**ADDITIONAL REFERENCES**

Apte, S. V., Gorokhovski, M. and Moin, P., "LES of atomizing spray with stochastic modeling of secondary breakup," International Journal of Multiphase Flow, Vol. 29, 2003, pp. 1503–1522.

Gordillo, J. M. and Fontelos, M. A., "Satellites in the inviscid breakup of bubbles," Physical Review Letters, Vol. 98, 2007, 144503.

Gordillo, J. M., "Axisymmetric bubble collapse in a quiescent liquid pool. I. Theory and numerical simulations," Physics of Fluids, Vol. 20, 2008, 112103.